\documentclass[aps,twocolumn]{revtex4}
\usepackage{ulem}
\usepackage{euscript}
\usepackage{epsfig}

\newcommand{\boldsymbol}[1]{\mbox{\boldmath $#1$}}

\begin{document}

\title{Atom-molecule equilibration in a degenerate Fermi gas with resonant interactions}
\author{J. E. Williams$^1$, T. Nikuni$^2$, N. Nygaard$^1$, C. W. Clark$^1$}

\affiliation{${}^1$Electron and Optical Physics Division, National Institute of Standards and Technology, Gaithersburg, Maryland 20899-8410}

\affiliation{${}^2$Department of Physics, Faculty of Science, Tokyo University of Science,1-3 Kagurazaka, Shinjuku-ku, Tokyo 162-8601, Japan.}

\begin{abstract}
We present a nonequilibrium kinetic theory describing atom-molecule population dynamics in a two-component Fermi gas with a Feshbach resonance. Key collision integrals emerge that govern the relaxation of the atom-molecule mixture to chemical and thermal equilibrium. Our focus is on the pseudogap regime where molecules form above the superfluid transition temperature. In this regime, we formulate a simple model for the atom-molecule population dynamics. The model predicts the saturation of molecule formation that has been observed in recent experiments, and indicates that a dramatic enhancement of the atom-molecule conversion efficiency occurs at low temperatures.
\end{abstract}

\maketitle

Over the past year, extraordinary progress has been made using a Feshbach resonance to control the atomic interactions in trapped Fermi gases \cite{MolAboveTc,MolBelowTc}. By adjusting an external magnetic field, the energy of a closed-channel bound state $\epsilon_{\rm{res}}(B)$ can be tuned relative to the $k=0$ threshold of a colliding pair of atoms \cite{Feshbach}. This has two dramatic effects on the system: atoms interact very strongly and molecules can be produced. 

Due to the resonant state coupling, the atom-atom scattering length $a(B)$ can be tuned to very large values, either positive or negative. One of the most stunning applications of this remarkable control of atomic interactions has been the recent achievement of pairing superfluidity in a dilute gas of fermionic atoms \cite{MolBelowTc}.  This remarkable tool allows one to continuously traverse the so-called BCS-BEC crossover region
\cite{BECBCS}.

In this paper we develop a kinetic theory for a resonantly coupled atom-molecule mixture in the pseudogap regime~\cite{Chen2004a} relevant to current experiments~\cite{MolAboveTc}. A key ingredient of our theory is a pair of newly identified collision integrals describing how the atom and molecule distributions come into chemical and thermal equilibrium. As a first application, we derive a non-interacting model for the population dynamics that predicts the saturation of molecule formation observed in recent experiments~\cite{MolAboveTc,MolBelowTc}.  It also predicts that the  efficiency of atom-to-molecule transfer increases as the temperature is lowered. 

During resonant collisions pairs of atoms form transient molecules, without the influence of the third atom required for recombination in nonresonant collisions \cite{Mies02a}. Such molecules are typically in a high-lying vibrational state that is very susceptible to collision-induced decay to a lower bound state. The energy released in such collisions quickly heats up the gas, a limiting factor that has frustrated attempts to achieve Bose-Einstein condensation of Feshbach molecules made from bosonic atoms~\cite{Mukaiyama03a}. 

For fermionic atoms such inelastic collisions are suppressed by approximately three orders of magnitude due to Fermi statistics \cite{threebody}, which results in lifetimes of the gas sample long enough that the molecules may relax to equilibrium in the trap. Consequently, an important question is how to properly describe the atom-molecule population dynamics in this regime \cite{Chin03a,Chin03b}. Theories based on a two-atom description, from which a two-level Landau-Zener treatment has emerged as a useful model \cite{Mies02a,Julienne04}, should only be valid in the regime $\tau_{\rm{sc}} \ll t \ll \tau_{\rm{col}}$, where $\tau_{\rm{sc}}\sim a/\bar v$ is the duration of a scattering event and $\tau_{\rm{col}}\sim 1/n\sigma \bar v$ is the average time between collisions. Here, $a$ is the s-wave scattering length, $\bar v$ is the mean velocity of an atom in the gas, and $\sigma$ is the collision cross-section.

In order to describe the current experimental regime, $\tau_{\rm{col}} \ll t$, we have derived coupled kinetic equations for the atom and molecule distributions above the superfluid transition temperature using the Keldysh nonequilibrium Green's function formalism \cite{Kadanoff1962a,Keldysh64a} .  Based on a {\it{selfconsistent}} diagramatic expansion, the atom and molecule self energies are treated on the same footing, leading to a renormalization of the atom and molecule mean field energy shifts. Before discussing the salient features of the theory, we first give a brief summary of the kinetic equations; a detailed derivation will appear in a future work \cite{Nikuni04}. We then derive a simplified model describing the atom-molecule population dynamics, which we use to understand recent experiments.

The starting point of our theory is the coupled boson-fermion Hamiltonian \cite{BECBCS}
\begin{eqnarray}
\hat{H} & = &  \int d \mathbf{r} \left [ \hat{\psi}_{\sigma}^{\dagger}(\mathbf{r})H_A(\mathbf{r}) \hat{\psi}_{\sigma}(\mathbf{r}) +   \hat{\phi}^{\dagger} (\mathbf{r}) 
H_M(\mathbf{r})\hat{\phi}(\mathbf{r}) \right ] \nonumber \\
&+& \kappa \int d \mathbf{r} \left [ \hat{\phi}^{\dagger}(\mathbf{r})
\hat{\psi}_{\downarrow}(\mathbf{r})\hat{\psi}_{\uparrow}(\mathbf{r}) + 
\hat{\psi}_{\uparrow}^{\dagger}(\mathbf{r})\hat{\psi}_{\downarrow}^{\dagger}(\mathbf{r})\hat{\phi}(\mathbf{r}) \right ]
\nonumber \\
&+& g_A \int d \mathbf{r} \hat{\psi}_{\uparrow}^{\dagger}(\mathbf{r})\hat{\psi}_{\downarrow}^{\dagger}(\mathbf{r})
\hat{\psi}_{\downarrow}(\mathbf{r})\hat{\psi}_{\uparrow}(\mathbf{r}) ,
\label{Hatommol}
\end{eqnarray}
where a sum over spin states $\sigma=\{\uparrow,\downarrow\}$ is implied in the first term. The atom  $\hat{\psi}_{\sigma}(\mathbf{r})$ and molecule $\hat{\phi}(\mathbf{r})$ field operators obey Fermi and Bose commutation relations, respectively. 
The single-particle hamiltonians for the atoms and molecules are given by
$H_A({\mathbf{r}}) = - ({\hbar^2}/2m)\nabla^2 + U_A(\bf{r})$ and $H_M({\mathbf{r}}) = - ({\hbar^2}/4m)\nabla^2 + U_M({\bf{r}}) + \epsilon_{\rm{res}}$,
where $U_A(\mathbf{r})$ and $U_M(\mathbf{r})$ are the external trapping potentials for atoms and molecules and $m$ is the atomic mass. In experiments, the energy $\epsilon_{\rm{res}}(B)$ of the resonant molecular state can be tuned by adjusting an external magnetic field $B$. The coupling to the resonant bound state is described by the two terms in the second line of (\ref{Hatommol}) with coupling constant $\kappa$; the purely open-channel part of atomic collisions is described by the last term with the interaction constant $g_A$. 

Starting with the many-body Hamiltonian (\ref{Hatommol}), we derive equations of motion for the
atomic and molecular distribution functions 
$f_\sigma({\mathbf{p}},{\mathbf{r}},t) = 
\int d {\mathbf{x}} e^{i {\mathbf{p}}\cdot {\mathbf{x}}/\hbar}  
\langle\hat{\psi}_{\sigma}^{\dagger}({\mathbf{r}}+{\mathbf{x}}/2)
\hat{\psi}_{\sigma}({\mathbf{r}}-{\mathbf{x}}/2) \rangle_t$ and
$f_M({\mathbf{p}},{\mathbf{r}},t) = \int d {\mathbf{x}}
e^{i {\mathbf{p}}\cdot {\mathbf{x}}/\hbar} 
\langle \hat{\phi}^{\dagger}({\mathbf{r}}+{\mathbf{x}}/2)
\hat{\phi}({\mathbf{r}}-{\mathbf{x}}/2) \rangle_t$,
where $\langle \cdots \rangle_t$ indicates a nonequilibrium statistical average. 
The steps leading to the kinetic equations are rather involved and are beyond the scope of this
letter \cite{Kadanoff1962a,Keldysh64a,Nikuni04}; here we state the results.
The kinetic equations describing the atom-molecule dynamics are
\begin{eqnarray}
\label{keA}
\frac{\partial f_\sigma({\mathbf{p}},{\mathbf{r}},t)}{\partial t} &-&
\left \{ \epsilon_\sigma({\mathbf{p}},{\mathbf{r}},t) , 
f_\sigma({\mathbf{p}},{\mathbf{r}},t)\right \} =
\mathcal{I}_\sigma + I_\sigma, \\
\frac{\partial f_M({\mathbf{p}},{\mathbf{r}},t)}{\partial t} &-&
\left \{ \epsilon_M({\mathbf{p}},{\mathbf{r}},t) , 
f_M({\mathbf{p}},{\mathbf{r}},t)\right \} =\mathcal{I}_M ,
\label{keM}
\end{eqnarray}
where $\{A,B\}\equiv{\boldsymbol{\nabla}}_{\mathbf{r}} A \cdot 
{\boldsymbol{\nabla}}_{\mathbf{p}} B - {\boldsymbol{\nabla}}_{\mathbf{p}} A
\cdot {\boldsymbol{\nabla}}_{\mathbf{r}} B $ is a Poisson bracket. 
The time-dependent renormalized energies for the atoms and molecules are given by (with the position dependence suppressed)
 $\epsilon_\sigma({\mathbf{p}}) =   \epsilon_A^{(0)}({\mathbf{p}}) +
\mathcal P\int d {\mathbf{p}}' g_{A,\sigma}({\mathbf{p}},{\mathbf{p}}')
f_{-\sigma}({\mathbf{p}}')/h^3 + \mathcal P \int d {\mathbf{p}}' 
g_{\kappa,\sigma}({\mathbf{p}},{\mathbf{p}}'  - {\mathbf{p}}) f_M({\mathbf{p}}')/h^3$
and $\epsilon_M({\mathbf{p}})  = \epsilon_M^{(0)}({\mathbf{p}}) + 
\mathcal P \int d {\mathbf{p}}' g_{\kappa,\uparrow}({\mathbf{p}}  - {\mathbf{p}}',{\mathbf{p}}')
[ f_\uparrow({\mathbf{p}}  -  {\mathbf{p}}')  + f_{\downarrow}({\mathbf{p}}')  -  1] $, where $\mathcal P$ indicates the principal value. The noninteracting parts of the energies are given by
$\epsilon_A^{(0)}({\mathbf{p}},{\mathbf{r}})={p^2}/{2m} + U_A({\mathbf{r}})$ and
$\epsilon_M^{(0)}({\mathbf{p}},{\mathbf{r}})={p^2}/{4m} + U_M({\mathbf{r}}) + \epsilon_{\rm{res}}$. The renormalized atom-atom interaction strength is 
 $  g_{A,\sigma}({\mathbf{p}},{\mathbf{p}}') \equiv g_A + g_{\kappa,\sigma}({\mathbf{p}},{\mathbf{p}}')$ where
$g_{\kappa,\sigma}({\mathbf{p}},{\mathbf{p}}') \equiv
\kappa^2 /[ \epsilon_\sigma({\mathbf{p}}) + \epsilon_{-\sigma}({\mathbf{p}}') 
- \epsilon_M({\mathbf{p}}+{\mathbf{p}}')]$.
The coupled integral equations for the renormalized energies must be solved
selfconsistently, since $\epsilon_\sigma({\mathbf{p}})$ and $\epsilon_M({\mathbf{p}})$
appear on both sides of the equations. 

The open-channel collision integral $I_\sigma$ takes the usual form for a quantum degenerate Fermi gas~\cite{Kadanoff1962a}, but with a renormalized momentum-dependent interaction strength~\cite{Nikuni04}. The collision integrals ${\mathcal{I}}_\sigma$ and ${\mathcal{I}}_M$ arise from the resonant state coupling and describe the formation and dissociation of molecules. They appear in both kinetic equations as reciprocal source/sink terms 
\begin{eqnarray}
\label{IcalA}
{\mathcal{I}}_\sigma &=& \frac{4 \pi^2 \, \kappa^2}{h^4 } \int 
d{\mathbf{p}}_2 \int d {\mathbf{p}}_3\delta({\bf p} + {\bf p}_2 - {\bf p}_3) \cr
&&\times\delta(\epsilon_\sigma({\mathbf{p}})+\epsilon_{-\sigma}({\mathbf{p}}_2)-\epsilon_M({\mathbf{p}}_3)) \nonumber \\
&\times& \Big \{ [1 - f_\sigma({\bf p})][1 - f_{-\sigma}({\bf p}_2)]f_M({\bf p}_3)  \cr
&-& f_\sigma({\bf p})f_{-\sigma}({\bf p}_2)[1+f_M({\bf p}_3)]
\Big \} .
\end{eqnarray}
The expression for the collision integral ${\mathcal{I}}_M$ in (\ref{keM}) can be obtained
as ${\mathcal{I}}_M=-{\mathcal{I}}_\sigma$ and by making the following 
substitutions on the r.h.s.  of (\ref{IcalA}):
$\{{\mathbf{p}}\rightarrow{\mathbf{p}}_1,{\mathbf{p}}_3\rightarrow{\mathbf{p}} \}$. 
From this relation it can be shown that the total atom population
$N_{\rm{tot}}=\int d {\mathbf{p}}\int d {\mathbf{r}} 
f_{\rm{tot}}({\mathbf{p}},{\mathbf{r}},t) / h^3$ is
conserved, where $f_{\rm{tot}}({\mathbf{p}},{\mathbf{r}},t)\equiv 
f_{\uparrow}({\mathbf{p}},{\mathbf{r}},t) + f_{\downarrow}({\mathbf{p}},{\mathbf{r}},t)
+2f_{M}({\mathbf{p}},{\mathbf{r}},t)$.

Equations (\ref{keA}) through (\ref{IcalA}), along with the supplementary definitions in the text, compose the nonequilibrium kinetic theory for a degenerate Fermi gas with resonant interactions. In deriving these equations we have made two important approximations about the molecules \cite{Nikuni04}: (i) our theory describes {\it{bare}} molecules since we neglect the molecular density renormalization and (ii) the finite molecular lifetime (given by the imaginary part of the molecule self-energy) is neglected in defining the quasi-particle energies $\epsilon_\sigma$ and $\epsilon_M$. The physics of the lifetime of the molecules (molecular dissociation) is, however, described by the collision integrals. The impact of approximation (ii) is that it ultimately leads to the appearance of the energy conserving delta functions in the collision integrals. We note that near the resonance, the molecules are actually dressed by the coupling to the atoms and are known to have a size of $a(B)/2$ \cite{Kohler03a,Duine03b,Bruun04b}, which is much larger than the size of a bare molecule. We expect our treatment to provide a good description when the molecules do not overlap spatially: $n [a(B)]^3 \ll 1$.  When $n [a(B)]^3 > 1$, the separation of timescales $\tau_{\rm{sc}}\ll\tau_{\rm{col}}\ll t$ assumption at the heart of our theory breaks down.

It is straightforward to show that the stationary solutions of (\ref{keA}) and (\ref{keM}) are Fermi and Bose distributions: $f_\sigma^0({\mathbf{p}},{\mathbf{r}})=\{e^{\beta[\epsilon_\sigma ({\mathbf{p}},{\mathbf{r}}) - \mu_\sigma]} + 1\}^{-1}$ and $f_M^0({\mathbf{p}},{\mathbf{r}})=\{e^{\beta[\epsilon_M ({\mathbf{p}},{\mathbf{r}}) - \mu_M]} - 1\}^{-1}$, where $\beta=1/k_{\rm{B}} T$. In order for ${\mathcal{I}}_\sigma$ and ${\mathcal{I}}_M$ to vanish, the chemical potential of the molecules must satisfy $\mu_M = \mu_\uparrow + \mu_\downarrow$. The collision integral ${\mathcal{I}}_M$ describes the process by which the molecular component comes into chemical and thermal equilibrium with the atoms, resulting in a Bose-Einstein distribution for the molecules. The molecular population depends strongly on the value and time rate of change of the resonant state energy $\epsilon_{\rm{res}}[B(t)]$. In our theory, molecules can only be produced when the renormalized resonance energy is tuned positive; on the negative detuning side of the resonance ($\epsilon_{\rm{res}}<0$), energy and momentum conservation cannot be satisfied, leading to ${\mathcal{I}}_\sigma={\mathcal{I}}_M=0$. 

Before giving an explicit application of the theory, it is important to comment on other relaxation processes that are {\it{not}} treated in our present theory. In Fig.~1 we give a summary of elastic and inelastic collisions that will determine how the atom-molecule mixture relaxes \cite{Chin03b}. The check marks indicate the processes described by our theory and the symbol $\gamma$ represents the relaxation rate for a given process. There are two other processes that play a role in forming molecules: two atoms form a molecule under the influence of a third atom ($A+A+A\leftrightarrow M+A$) or molecule ($A+A+M\leftrightarrow M+M$). As discussed above, on the negative side of the resonance $\gamma_{AA}=0$; however, in general $\gamma_{AAA}$ and $\gamma_{AAM}$ {\it{do not}} vanish. Our kinetic equations are therefore a good description when $\gamma_{AAA}$ and $\gamma_{AAM}$ are much smaller than the other relaxation rates. We finally note that  inelastic collisions, such as those listed in Fig.~1, are assumed to be negligible.
\begin{figure}
  \centerline{\epsfig{file=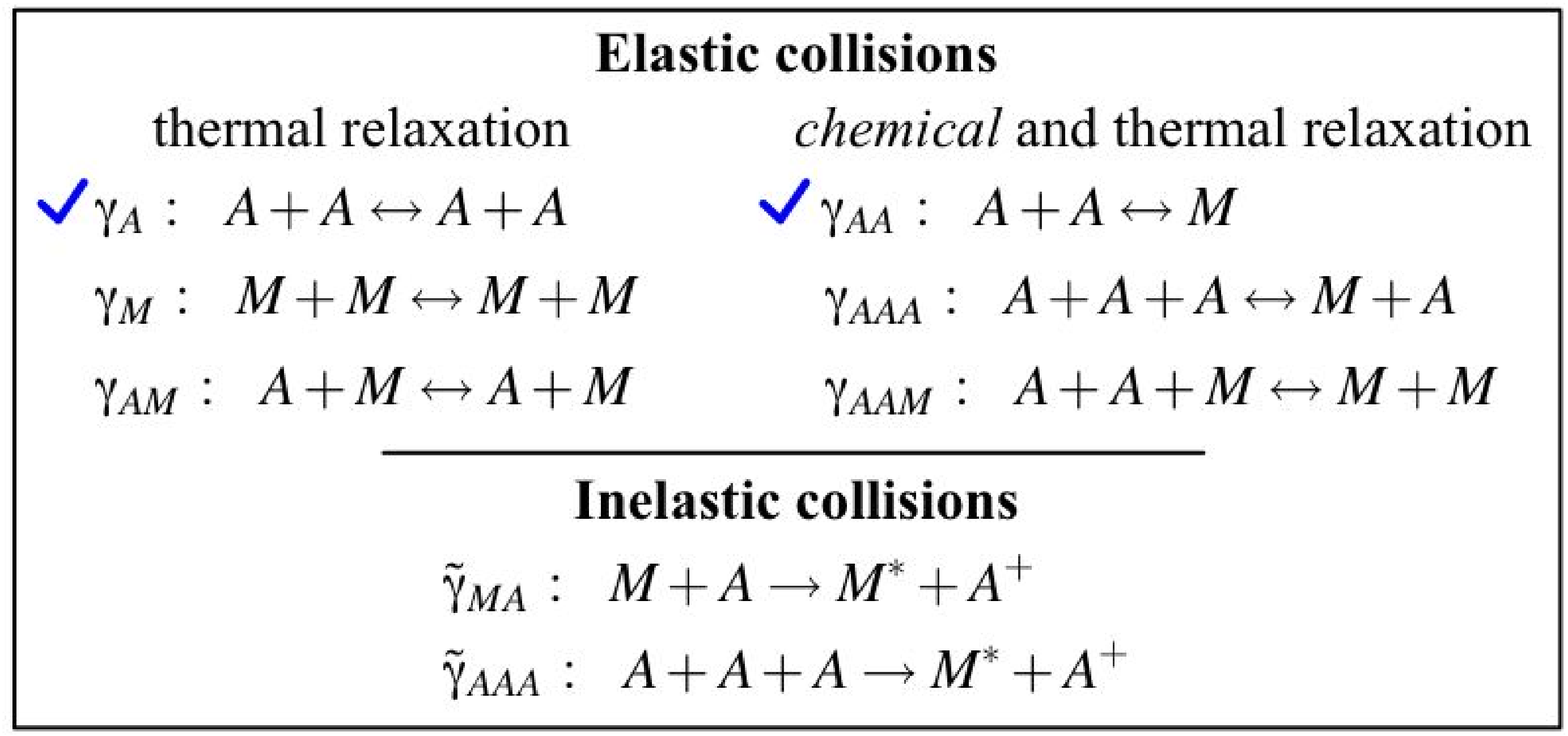,width=3.25in}}
\caption{Summary of different relaxation processes. The letter $A$ signifies an atom in either spin component, $M$ refers to the resonant molecule, $M^*$ describes a more deeply bound molecule, and "$+$" denotes that the particle carries away an energy far in excess of the thermal energy $k_{\rm{B}} T$. The check marks designate the processes that are treated in this letter.}
\end{figure}

We now derive a simplified model to describe the atom-molecule population dynamics. We assume an equal mixture of spin states $f_\uparrow=f_\downarrow\equiv f_A$ and take $U_M({\bf{r}})=2U_A({\bf{r}})$. Our goal is to obtain a coupled set of dynamical equations describing the populations $N_i(t)$ and energies $E_i(t)$ of the atom and molecule distributions, which are defined by $N_i(t)=\int d \mathbf{r} \int  d \mathbf{p}
f_i(\mathbf{p},\mathbf{r},t)/h^3$ and $E_i(t)=\int d \mathbf{r} \int d \mathbf{p}\epsilon_i(\mathbf{p},\mathbf{r},t)
f_i(\mathbf{p},\mathbf{r},t)/h^3$. The subscript $i$ denotes atoms or molecules $i\in\{A,M\}$. 

Equations of motion for $N_i(t)$ and $E_i(t)$ are obtained by differentiating with respect to time and substituting the kinetic equations from (\ref{keA}) and (\ref{keM}).
We make the following three simplifications in order to obtain a model description:
(a) We neglect the renormalized mean field contribution to the energies,
and everywhere make the substitution
$\epsilon_i (\mathbf{p},\mathbf{r},t) \rightarrow \epsilon^{(0)}_i (\mathbf{p},\mathbf{r})$.  (b) For temperatures not too deep into the quantum degenerate regime (e.g. $T/T_{\rm{F}}>0.5$ \cite{DeMarco01a}), a classical gas provides a reasonably good description. We therefore take $(1\pm f_i)\rightarrow 1$ in the collision integrals. (c) We finally assume Maxwell-Boltzmann forms for the distributions, but allow the temperatures  and chemical potentials to vary in time: $f_A(\mathbf{p},\mathbf{r},t)=\exp\{-[\epsilon_A^{(0)}(\mathbf{p},\mathbf{r}) - \mu_A(t)]/k_{\rm{B}} T_A(t)\}$ and 
$f_M(\mathbf{p},\mathbf{r},t) = \exp\{ -[\epsilon_M^{(0)} (\mathbf{p},\mathbf{r}) - \mu_M(t)]/k_{\rm{B}} T_M(t) \}$. 

It is now straightforward to obtain a closed set of coupled differential equations for the populations $N_i(t)$
and energies $E_i(t)$. The atom equations are
\begin{eqnarray}
\label{NAdot}
\frac{d N_A}{dt} &=& \gamma_\kappa(\epsilon_{\rm{res}}) \Big [N_M - N_Ae^{-( \epsilon_{\rm{res}}- \mu_A)/k_{\rm{B}} T_A} \Big ] ,\\
\label{EAdot}
\frac{d E_A}{dt} &=& \frac{\gamma_\kappa(\epsilon_{\rm{res}})}{2} \Big [E_M - (E_A+\epsilon_{\rm{res}}N_A)e^{-( \epsilon_{\rm{res}}- \mu_A)/k_{\rm{B}} T_A} \Big ].
\nonumber \\
\end{eqnarray}
The variables $\mu_i(t)$ and $T_i(t)$ are related to the system variables $\{N_A(t),N_M(t),E_A(t),E_M(t)\}$ using the definitions of $N_i(t)$ and $E_i(t)$ given above. Due to number conservation, the molecule population is given by $N_M(t)=N_{\rm{tot}}/2 - N_A(t)$. The equation for the molecule energy is $d E_M(t)/dt = \dot \epsilon_{\rm{res}} N_M(t) - 2 d E_A(t)/dt$, where $\dot \epsilon_{\rm{res}}$ is the time rate of change of the resonant state energy. The total energy $E_{\rm{tot}}(t)=2E_A(t)+E_M(t)$ varies as $dE_{\rm{tot}}/dt=\dot \epsilon_{\rm{res}}N_M(t)$. It can be shown that the r.h.s. of (\ref{NAdot}) and (\ref{EAdot}) vanish when $\mu_M=2\mu_A$ and $T_M = T_A$.

The relaxation rate $\gamma_\kappa(\epsilon)$  is given by
\begin{equation}
\label{gamma}
\gamma_\kappa(\epsilon) = \frac{\kappa^2 m^{3/2}}{2\pi \hbar^4} \sqrt{\epsilon} \,\Theta(\epsilon),
\end{equation}
which has the form of a Fermi's golden rule \cite{Mukaiyama03a}. The Heaviside step function $\Theta(\epsilon)$ signifies that for negative detunings, the exchange of atoms and molecules ceases. This rate determines whether a sweep of $\epsilon_{\rm{res}}(t)$ is adiabatic in the sense that the entropy is conserved.  Using dimensional analysis, we find a good measure of adiabaticity in our calculations is  $\alpha/\dot \epsilon_{\rm{res}}$, where $\alpha= k_{\rm{B}} T \,\gamma_\kappa(k_{\rm{B}} T)$. The process is adiabatic if $\alpha/\dot \epsilon_{\rm{res}} \gg 1$. 

To express this inequality in terms of physical parameters, we approximate the coupling strength by $\kappa^2\approx 2\pi \hbar^2 a_{\rm{bg}} \Delta\mu \, \Delta B/m$ \cite{Duine03b}, where $a_{\rm{bg}}$ is the background scattering length, $\Delta\mu$ is the difference in magnetic moments of the open and closed channels, and $\Delta B$ is the resonance width. Our theory does not provide the magnetic field dependence of $\epsilon_{\rm{res}}(B)$, so we make the simple approximation
$\dot \epsilon_{\rm{res}}\approx\Delta\mu \dot B$. With these estimates, we obtain $\alpha/\dot \epsilon_{\rm{res}}\approx(2\pi)^{3/2}(\hbar a_{\rm{bg}}/m\lambda_{\rm{th}}^3) \Delta B/\dot B$, where $\lambda_{\rm{th}}=h/\sqrt{2\pi m k_{\rm{B}} T}$ is the thermal deBroglie wavelength. This expression resembles the Landau-Zener adiabaticity requirement  \cite{Mies02a}, but with the volume factor $\lambda_{\rm{th}}^3$ appearing for a thermalized gas at finite temperatures.

 We solve the coupled population and energy equations numerically using physical parameters for the recent experiments by Regal {\it{et al}}. \cite{MolAboveTc,MolBelowTc} with ${}^{40}$K at the resonance $B_0=20.21$ mT, for which $\Delta B = 0.78$ mT and  $a_{\rm{bg}}=174 a_0$, where $a_0$ is the Bohr radius. We take $N_{\rm{tot}}=10^6$ and trap frequencies $\nu_x=\nu_y=290$ Hz and $\nu_z = \nu_x/80$. The Fermi temperature is $T_{\rm{F}} = 0.47 \mu$K. We present results for $T/T_{\rm{F}}=0.5$, which is in the range $0.1 < T/T_{\rm{F}} < 2.0$ that applies to recent experiments on molecule formation~\cite{MolAboveTc}. Taking $\Delta \mu\approx2\mu_{\rm{B}}$, the rate constant is $\alpha/\mu_{\rm{B}}=1.9$ T/s, where $\mu_{\rm{B}}$ is the Bohr magneton. 
 
The resonance energy is ramped down linearly as $\epsilon_{\rm{res}}(t)=\epsilon_{\rm{res}}(t=0) - \dot \epsilon_{\rm{res}} t$ and stopped on the negative detuning side. In Fig.~2, we plot the molecule fraction $\eta_M\equiv 2N_M/N_{\rm{tot}}$ versus $\epsilon_{\rm{res}}(t)$ for two different sweep rates: $\alpha/\dot \epsilon_{\rm{res}}=5$ (red dashed line) and $\alpha/\dot \epsilon_{\rm{res}}=0.2$ (blue dot-dashed line). For comparison, we plot the equilibrium solution for a classical ideal gas mixture of atoms and molecules along a line of constant entropy, given by the solid black line. When the adiabaticity criterion is satisfied $\alpha/\dot \epsilon_{\rm{res}}\gg1$, the sweep achieves the maximum transfer allowed by following along a line of constant entropy, before entering the negative side of the resonance where the atom-molecule transfer stops. Recall our
assumption that the rates $\gamma_{AAA}$ and $\gamma_{AAM}$ appearing in Fig.~1 are much smaller than $\gamma_{AA} \sim \gamma_\kappa$. If the ramp is carried out slowly compared to these 3-body relaxation rates, the system should follow the solid black curve on the negative side, which approaches unity.
\begin{figure}
  \centerline{\epsfig{file=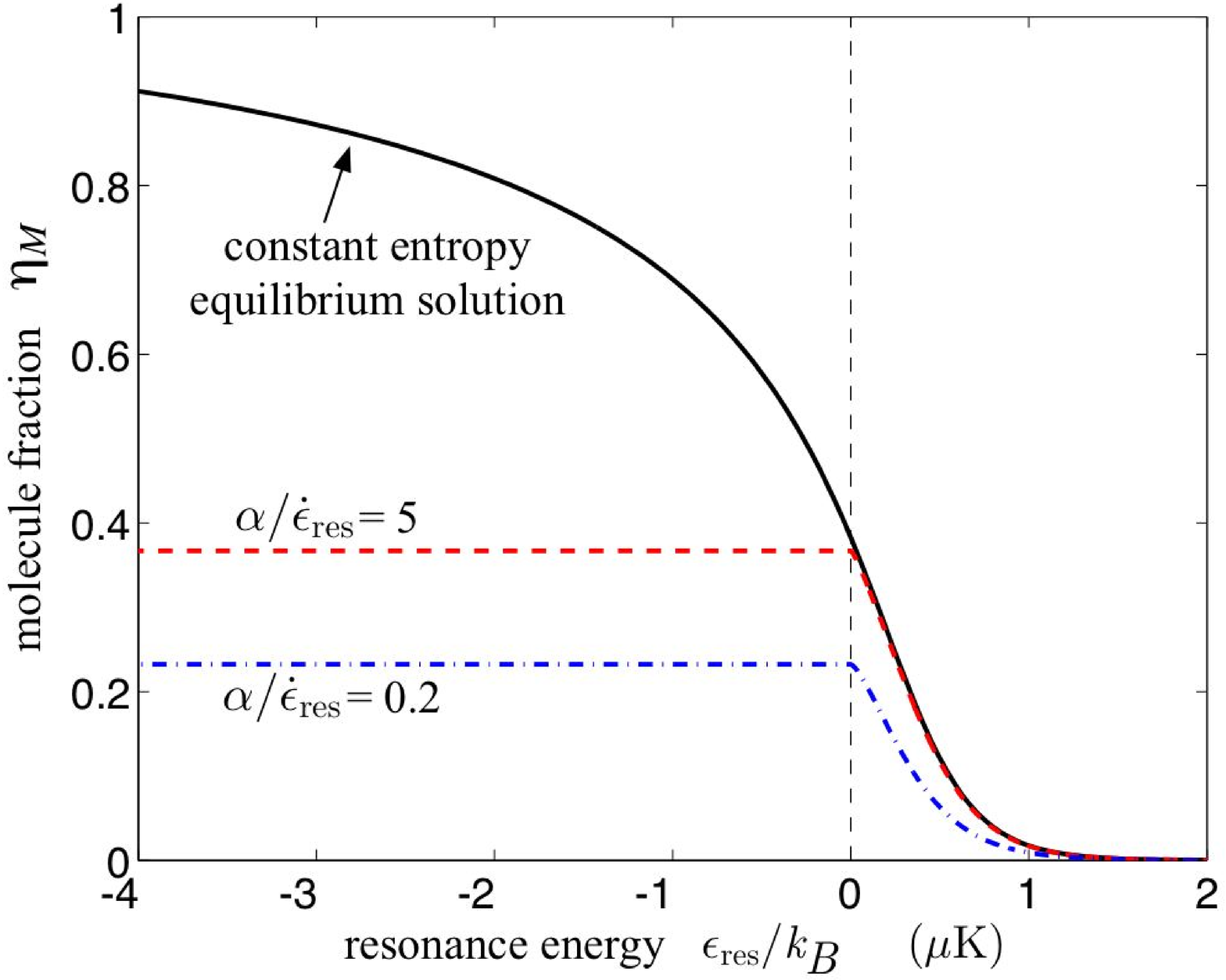,width=3in}}
\caption{Molecule fraction versus the resonance energy for two different sweep rates. The red dashed and blue dot-dashed curves are obtained by integrating the equations of motion for $\{N_A(t),N_M(t),E_A(T),E_M(t)\}$ with a linear ramp of $\epsilon_{\rm{res}}(t)$, starting with $\epsilon_{\rm{res}}(t=0)/k_{\rm{B}}=2 \, \mu$K. The black solid curve is obtained by calculating the equilibrium solution for a classical ideal gas mixture of atoms and molecules. The initial temperature is $T(t=0)/T_{\rm{F}}=0.5$.}
\end{figure}

In Fig.~3 we plot the final atom fraction $\eta_A=1-\eta_M$ versus the inverse sweep rate. An important result is that it is impossible to achieve a perfect transfer of all the atoms into molecules at finite temperatures, even when the sweep is adiabatic; one can only achieve the limiting value $\min\{\eta_A\}$. This saturation of the transfer efficiency is observed in experiments \cite{MolAboveTc,MolBelowTc}. The value of $\min\{\eta_A\}$ is strongly dependent on temperature, as seen in the inset of Fig.~3 where we plot $\min\{\eta_A\}$ versus $T$. Our model predicts that the transfer efficiency ($1-\min\{\eta_A\}$) increases as $T$ decreases. We note that for temperatures below $T/T_{\rm{F}} \approx 0.5$, the effects of quantum statistics, which are neglected in our simple model, will become important.

We have presented kinetic equations (\ref{keA}) and (\ref{keM}) describing a resonantly coupled atom-molecule mixture in the pseudogap regime above the superfluid transition. As a first application, we derived a non-interacting model for the population dynamics, similar to the toy model presented in Ref.~\cite{Williams2004b} for {\it{equilibrium}} properties. The saturation of molecule formation predicted by our model was observed in recent experiments above~\cite{MolAboveTc} and below~\cite{MolBelowTc} the superfluid transition temperature. We also predict that the transfer efficiency increases as the temperature is lowered. The kinetic equations provide a foundation for many future applications of dynamical effects, such as exploring the collisionless to hydrodynamic crossover of the damped collective excitations. 
\begin{figure}
  \centerline{\epsfig{file=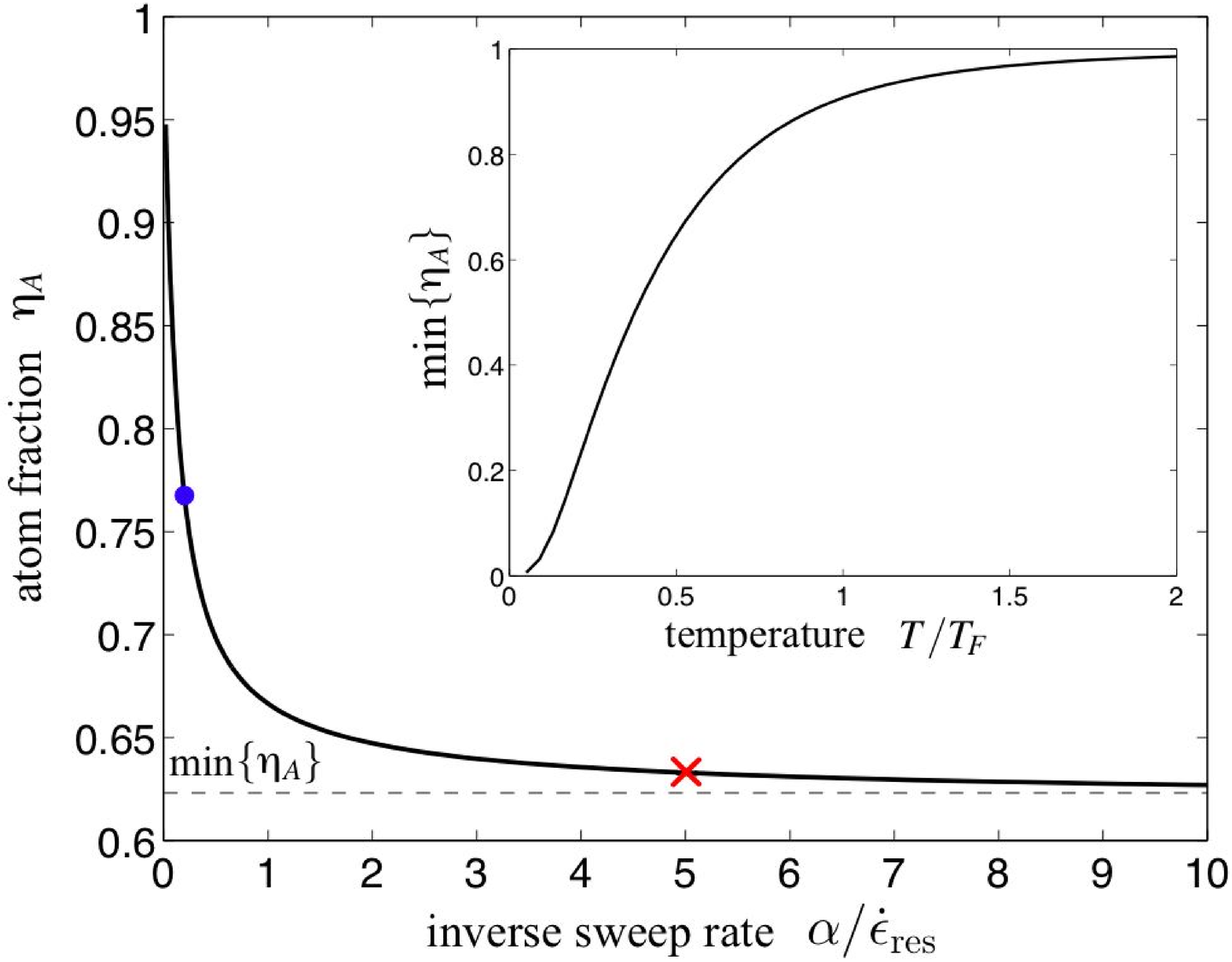,width=3in}}
\caption{Atom fraction versus the inverse sweep rate. The blue circle and red cross correspond to the blue dot-dashed and red dashed curves shown in Fig.~2, respectively. The inset shows how $\min\{\eta_A\}$ decreases with decreasing temperature.}
\end{figure}

We acknowledge J. Cooper, A. Griffin, M. Holland, P. Julienne, and E. Tiesinga for helpful discussions.


\begin{thebibliography}{10}

\bibitem{MolAboveTc}
C.~A. Regal {\it et~al.}, Nature {\bf 424},  47  (2003); K. Strecker, G. Partridge, and R. Hulet, Phys. Rev. Lett. {\bf 91},  080406  (2003); J. Cubizolles {\it et~al.}, {\it{ibid}}. {\bf 91},  240401  (2003); S. Jochim {\it et~al.}, {\it{ibid}}. {\bf 91},  240402  (2003).

\bibitem{MolBelowTc}
M. Greiner, C. Regal, and D. Jin, Nature {\bf 426},  537  (2003); S. Jochim {\it et~al.}, Science {\bf 302},  2101  (2003); M. Zwierlein {\it et~al.}, Phys. Rev. Lett. {\bf 91},  250401  (2003); C.~A. Regal, M. Greiner, and D.~S. Jin, {\it{ibid}}. {\bf 92},  040403  (2004); M. Bartenstein {\it et~al.}, {\it{ibid}}. {\bf 92},  120401  (2004); M. Zwierlein {\it et~al.}, {\it{ibid}}. {\bf{92}}, 120403 (2004).

\bibitem{Feshbach}
W. Stwalley, Phys. Rev. Lett. {\bf 37},  1628  (1976); E. Tiesinga, B. Verhaar, and H. Stoof, Phys. Rev. A {\bf 47},  4114  (1993).

\bibitem{BECBCS}
M. Holland {\it et~al.}, Phys. Rev. Lett. {\bf 87},  120406  (2001); E. Timmermans {\it et~al.}, Phys. Lett. A {\bf 285},  228  (2001); Y. Ohashi and A. Griffin, Phys. Rev. Lett. {\bf 89},  130402  (2002).

\bibitem{Chen2004a}
Q. Chen {\it et~al.}, cond-mat/0404274.

\bibitem{Mies02a}
F. Mies, E. Tiesinga, and P. Julienne, Phys. Rev. A {\bf 61},  022721  (2002).

\bibitem{Mukaiyama03a}
T. Mukaiyama {\it et~al.}, Phys. Rev. Lett. {\bf{92}}, 180402 (2004).

\bibitem{threebody}
B. Esry, C. Greene, and H. Suno, Phys. Rev. A {\bf 65},  010705  (2001); D. Petrov, C. Salomon, and G. Shlyapnikov, cond-mat/0309010; D. Petrov, Phys. Rev. A {\bf 67},  010703  (2003).

\bibitem{Chin03a}
C. Chin {\it et~al.}, Phys. Rev. Lett. {\bf 90},  033201  (2003).

\bibitem{Chin03b}
C. Chin and R. Grimm, Phys. Rev. A {\bf 69},  033612  (2004).

\bibitem{Julienne04}
P. Julienne, E. Tiesinga, and T. K\"ohler, cond-mat/0312492  .

\bibitem{Kadanoff1962a}
L.~P. Kadanoff and G. Baym, {\it Quantum Statistical Mechanics} (W. A.
  Benjamin, New York, 1962).

\bibitem{Keldysh64a}
L. Keldysh, Zh. Eksp. Teor. Fiz. {\bf 47},  1515  (1964).

\bibitem{Nikuni04}
T. Nikuni {\it et~al.}, unpublished  .

\bibitem{Kohler03a}
T. K\"ohler {\it et~al.}, Phys. Rev. Lett. {\bf 91},  230401  (2003).

\bibitem{Duine03b}
R. Duine and H. Stoof, J. Opt. B: Quantum Semiclass. Opt. {\bf 5},  S212 (2003).

\bibitem{Bruun04b}
G.M. Bruun and C.J. Pethick, Phys. Rev. Lett. {\bf 92},  140404 (2004).

\bibitem{DeMarco01a}
B. DeMarco, S. Papp, and D. Jin, Phys. Rev. Lett. {\bf 86},  5409  (2001).

\bibitem{Williams2004b}
J. E. Williams, N. Nygaard and C. W. Clark, unpublished.

\end{thebibliography}
 \end{document}